\documentclass[superscriptaddress,amsmath,amssymb,aps,prx,reprint]{revtex4-2}

\usepackage{graphicx}
\usepackage{dcolumn}
\usepackage{bm}
\usepackage{hyperref}
\usepackage{notes2bib}
\usepackage{siunitx}
\usepackage{footnotebackref}
\usepackage[usenames,dvipsnames]{xcolor}

\usepackage{lineno}

\definecolor{PR}{rgb}{0.57, 0.36, 0.51}
\definecolor{todo}{rgb}{1, 0.75, 0.0}

\begin{document}


\title{Fabry-Pérot cavities and quantum dot formation at gate-defined interfaces in twisted double bilayer graphene}

\author{El\'ias Portolés$^{\dagger}$}
\email{eliaspo@phys.ethz.ch}
\author{Giulia Zheng}
\email{These authors contributed equally}
\author{Folkert K. de Vries}
\affiliation{Solid State Physics Laboratory, ETH Zürich,~CH-8093~Zürich, Switzerland}
\author{Jihang Zhu}
\affiliation{Department of Physics, University of Texas at Austin, Austin, Texas 78712, USA}
\author{Petar Tomić}
\affiliation{Solid State Physics Laboratory, ETH Zürich,~CH-8093~Zürich, Switzerland}
\author{Takashi Taniguchi}
\affiliation{International Center for Materials Nanoarchitectonics,
National Institute for Materials Science,  1-1 Namiki, Tsukuba 305-0044, Japan}
\author{Kenji Watanabe}
\affiliation{Research Center for Functional Materials,
National Institute for Materials Science, 1-1 Namiki, Tsukuba 305-0044, Japan}
\author{Allan H. MacDonald}
\affiliation{Department of Physics, University of Texas at Austin, Austin, Texas 78712, USA}
\author{Klaus Ensslin}
\author{Thomas Ihn}
\author{Peter Rickhaus}
\affiliation{Solid State Physics Laboratory, ETH Zürich,~CH-8093~Zürich, Switzerland}

\begin{abstract}
The rich and electrostatically tunable phase diagram exhibited by moiré materials has made them a suitable platform for hosting single material multi-purpose devices. To engineer such devices, understanding electronic transport and localization across electrostatically defined interfaces is of fundamental importance. Little is known, however, about how the interplay between the band structure originating from the moiré lattice and electric potential gradients affects electronic confinement. Here, we electrostatically define a cavity across a twisted double bilayer graphene sample. We observe two kinds of Fabry-Pérot oscillations. The first, independent of charge polarity, stems from confinement of electrons between dispersive-band/flat-band interfaces. The second arises from junctions between regions tuned into different flat bands. When tuning the out-of-plane electric field across the device, we observe Coulomb blockade resonances in transport, an indication of strong electronic confinement. From the gate, magnetic field and source-drain voltage dependence of the resonances, we conclude that quantum dots form at the interfaces of the Fabry-Pérot cavity. Our results constitute a first step towards better understanding interfacial phenomena in single crystal moiré devices.
\end{abstract}

\maketitle
\footnotetext{These authors contributed equally to this work}
\section{Introduction}
The interest in moiré crystals, sparked by the advent of magic-angle twisted bilayer graphene~\cite{Cao2018_1,Cao2018_2}, has also drawn attention to other material combinations. One of those combinations is Twisted Double Bilayer Graphene (TDBG).
This material consists of two Bernal-stacked graphene bilayers that are stacked on top of each other with a certain twist angle. It combines the strong interactions originating from the flatness of the bands and the out-of-plane electric field (displacement field) tunability of Bernal-stacked bilayer graphene~\cite{McCann_2013, Bistritzer12233}. The band structure of small-angle TDBG shows flat and dispersive bands~\cite{Cao2020, Shen2020}. Tuning the charge carrier density can change the nature of the electronic phases to, for example, correlated insulators or ferromagnets~\cite{Liu2020, Burg2019, Cao2020, Shen2020}.
In order to explore nanostructures in moiré materials~\cite{deVries2021, Rodan-Legrain2021}, understanding transport across gate-defined interfaces is of importance.

Interfaces between different materials have proven to be ideal hosts for a plethora of physical phenomena~\cite{Kroemer2001}, in particular since the development of epitaxially grown, atomically sharp vertical interfaces. However, the unavoidable lattice mismatch between crystals often limits its quality. Moiré materials constitute a platform in which single crystal interfaces can be engineered, avoiding the mismatch. Lateral interfaces in such materials might, like their vertical counterparts, exhibit rich physics, with the advantage of being tunable. Because of the electric field gradient, the \textit{in situ} tunability of gate-defined interfaces comes at the expense of atomic sharpness. Combining two interfaces, an electronic cavity can be formed. Fabry-Pérot (FP) oscillations are one of the most studied phenomena arising from transport through a cavity ~\cite{Liang2001, Young2009, Rickhaus2013}. Apart from being of interest as a physical phenomenon on their own, the oscillations are also a useful tool for characterizing the charge carriers traversing the cavity.


In analogy with a mirror-defined optical cavity, an electronic FP interferometer is formed by partially transmitting potential steps. In a one-dimensional system, where the mean free path is much longer than the cavity length, successive partial reflections of electronic waves at the steps lead to interference. In two dimensions, electron waves can enter the cavity at different incident angles, leading to different effective wavelengths normal to the cavity boundary. The resulting averaging of a large range of interference periods leads to a vanishing interference pattern. Therefore, in order to observe a coherent pattern, angle selectivity of the incident angle is required. This can be provided by different effects arising from a difference in band structure between leads and cavity. Examples include electron-hole scattering in band inverted materials~\cite{Karalic2020}, Klein tunnelling for monolayer graphene~\cite{Katsnelson2006} and suppression of anti-Klein tunnelling for bilayer graphene~\cite{Varlet2014}. 

Stronger electronic confinement than in cavities with semi-transparent mirrors has been observed in Bernal-stacked bilayer graphene in the form of quantum dots~\cite{Eich2018, Banszerus2018}. There, dots are formed by an interplay between displacement field, band gaps and alternating charge-carrier polarities. Compared to bilayer graphene, in TDBG the moiré potential originating from the twist is added. It is addressed in this manuscript, how this combines with the previously mentioned elements contributing to the formation of dots across smooth barriers.

Here, we study electronic transport across an electrostatically defined cavity in a TDBG sample with a twist angle of $1.07^{\circ}$. We form two types of FP cavities. First, we investigate  dispersive/flat band interfaces both with same and opposite charge polarity of carriers. Second, we show the formation of cavities between different flat bands.
At large displacement fields, we observe Coulomb blockade resonances in the conductance. We characterize the apparent quantum dots in a magnetic field and extract an upper bound for their size, an estimate of the charge-carrier's g-factor and their charging energy. From the data we trace back the origin of the dots to strong electron confinement at the interface between cavity and leads. This might originate from an interplay between the density gradient and the moiré crystal.

\begin{figure}[p!]
\includegraphics[width=0.5\textwidth]{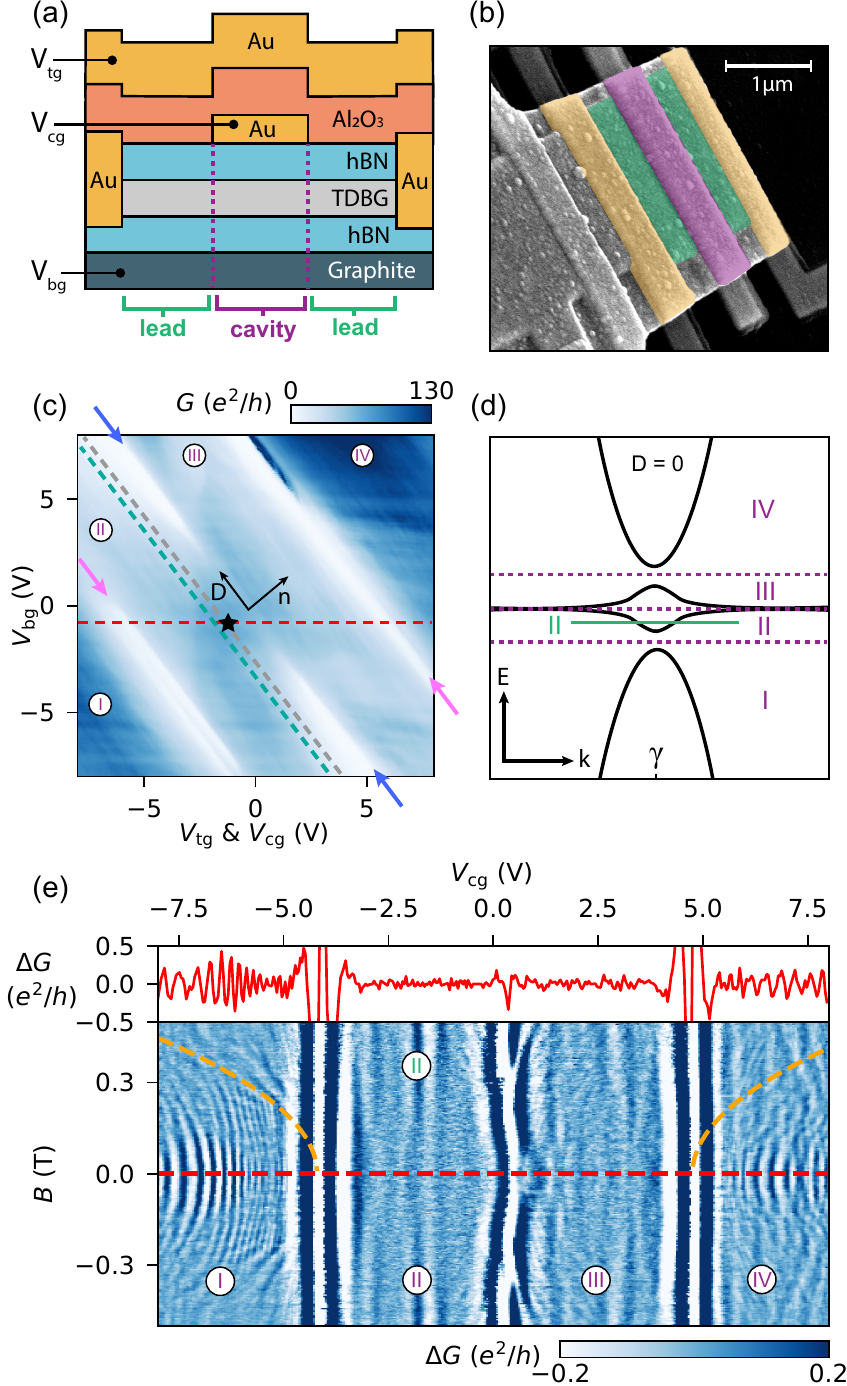}
\caption{(a) Side view schematic of the device. Gold electrodes are labelled 'Au'. $\mathrm{Al_2O_3}$ represents the aluminum oxide. hBN stands for hexagonal Boron Nitride and TDBG for Twisted Double Bilayer Graphene. The central gate (biased by $V_{\mathrm{cg}}$) allows to define a cavity. (b) False colored scanning electron microscopy image of a representative device. Orange highlights the contacts, green and purple correspond to leads and central gate, respectively. (c) Conductance as a function of top and bottom gates. Pink (blue) arrows point to gaps at full filling (charge neutrality point). The gaps divide the map in four regions labelled with roman numbers. The star indicates the value of $V_{\mathrm{tg}}$ and $V_{\mathrm{bg}}$ in Fig~\ref{fig:1}(e), while the red dashed line indicates the sweep range of $V_{\mathrm{cg}}$ in the same figure. The grey (turquoise) dashed line indicates the ranges of $V_{\mathrm{tg}}$ and $V_{\mathrm{bg}}$ in Fig.~\ref{fig:2}(a) (Fig.~\ref{fig:2}(d)). (d) Schematics of the band structure for $D=0$. Purple dashed lines delimit bands corresponding to regions I to IV in Fig~\ref{fig:1}(e). The green line indicates the Fermi energy in the leads in Fig~\ref{fig:1}(e). For the band structure at finite $D$ see Fig.~\ref{fig:3}(a). (e) Upper panel: relative conductance as a function of $V_{\mathrm{cg}}$ at $D_{\mathrm{l}} = \SI{0}{V/nm}$ and $n_{\mathrm{l}} = \SI{-1e12}{cm^{-2}}$. Lower panel: same trace as a function of magnetic field $B$. Dashed red line corresponds to data in the upper panel. Dashed yellow lines indicate the $B$ values at which the cyclotron radius equals the length of the cavity.}
\label{fig:1}
\end{figure}

\section{Fabry-Pérot cavity}
We fabricate our stack [see Fig.~\ref{fig:1}(a)] by first cutting a bilayer graphene flake into two parts with a needle (tungsten, $\SI{2}{\mu m}$ tip diameter). We successively pick up flakes to obtain a $1.07^{\circ}$ twisted graphene structure, encapsulated by two layers of hexagonal Boron Nitride (hBN) and a graphite back gate~\cite{Zomer2014, Kim2016} (See Supplemental Material for details about how we determine the twist angle). We selectively etch the structure and evaporate gold (Au) ohmic-contacts of $\SI{110}{nm}$ thickness. Another evaporation of gold on top of hBN follows, this time for the gates, which define the FP cavity of a length of $\SI{400}{nm}$. We then define the mesa by etching and deposit a $\SI{30}{nm}$ thick $\mathrm{Al_2O_3}$ layer by using atomic layer deposition (ALD). Finally, we evaporate another $\SI{110}{nm}$ Au layer to define the global top gate which is used to electrostatically bias the sample. Figure~\ref{fig:1}(b) shows a scanning electron microscope top-view of a representative device.

\begin{figure*}[t]
\centering
\includegraphics[width=1\textwidth]{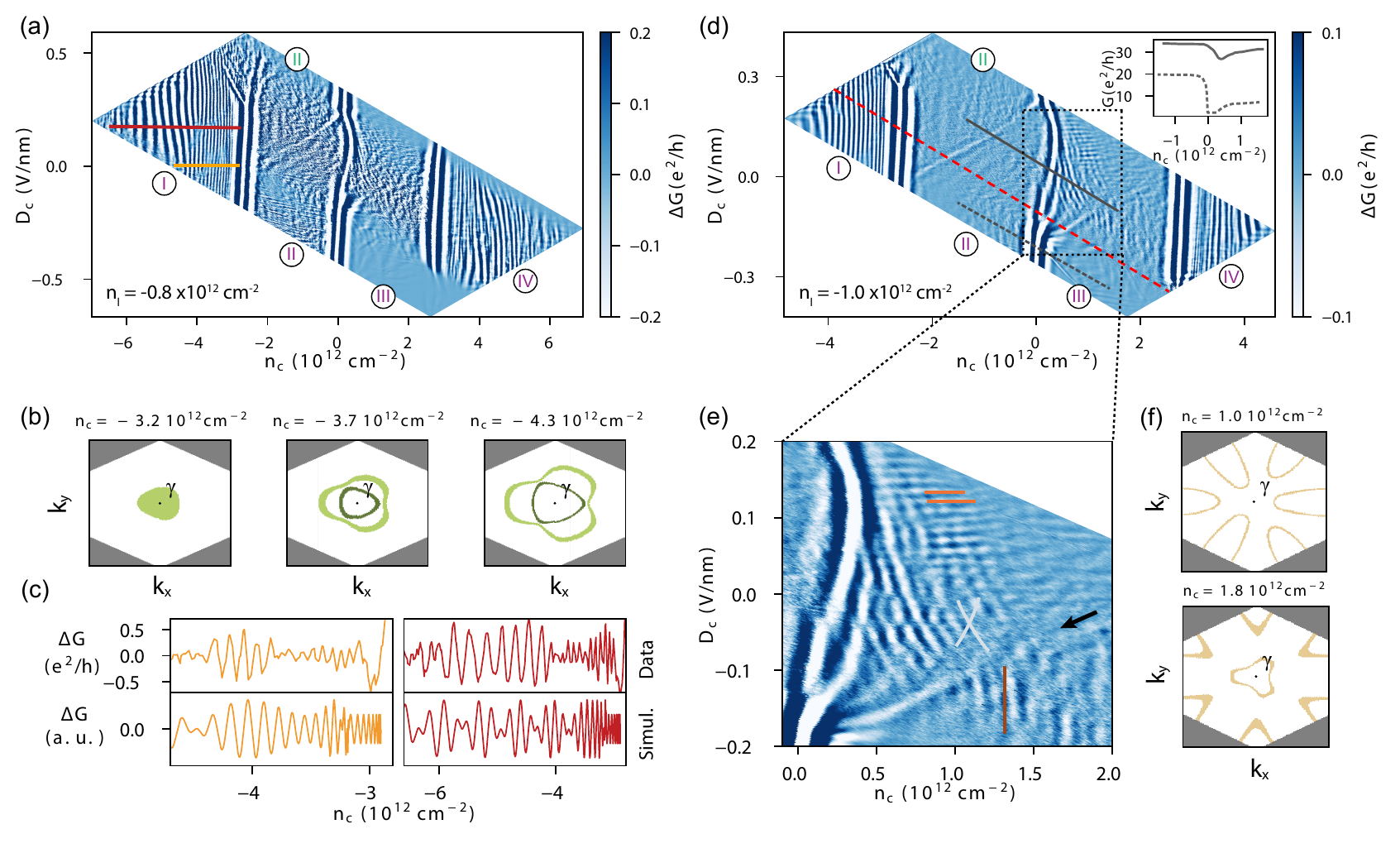}
\caption{(a) Relative conductance (see Supplemental Material) as a function of displacement field and electron density in the cavity. The density in the leads is $n_{\mathrm{l}} = \SI{-0.8e12}{cm^{-2}}$. Roman numbers indicate regions in Fig.~\ref{fig:1}(d). (b) Constant energy contours in the first Brillouin zone for $n_{\mathrm{c}} = \SI{-3.2e12}{cm^{-2}}$, $n_{\mathrm{c}} = \SI{-3.7e12}{cm^{-2}}$ and $n_{\mathrm{c}} = \SI{-4.3e12}{cm^{-2}}$. The light (dark) green color refers to the first (second) dispersive band. (c) The top panels show measurement of conductance versus density in the cavity in region I of Fig.~\ref{fig:2}(a) for $D_{\mathrm{c}}= \SI{0}{V/nm}$ and $D_{\mathrm{c}}=\SI{0.17}{V/nm}$  (red and orange lines in Fig.~\ref{fig:2}(a)). The bottom panels are the corresponding simulations. (d) Relative conductance as a function of displacement field and electron density in the cavity. The density in the leads is $n_{\mathrm{l}} = \SI{-1e12}{cm^{-2}}$. The inset shows conductance along the grey solid and dashed lines at $D_{\mathrm{l}} = \SI{0.15}{V/nm}$ and $D_{\mathrm{l}} = \SI{-0.1}{V/nm}$ respectively. The red dashed line in this figure corresponds to the red dashed line in Figs.~\ref{fig:1}(c) and \ref{fig:1}(e), at $D_{\mathrm{l}} = \SI{0}{V/nm}$. Here the sweeping range of $V_\mathrm{c}$ is smaller. (e) Zoom into the region in the black dotted square of Fig.~\ref{fig:2}(d). Brown and white lines indicate oscillations which are density dependent, while the orange horizontal lines are only displacement field dependent. (f) Cuts of the Fermi surfaces of the electron flat conduction band in the first Brillouin zone for $n_{\mathrm{c}} = \SI{1e12}{cm^{-2}}$ and  $n_{\mathrm{c}}= \SI{2e12}{cm^{-2}}$. The different topologies of the two panels indicate the crossing of a Lifshitz transition.}
\label{fig:2}
\end{figure*}

\begin{figure*}[t]
\centering
\includegraphics[width=1\textwidth]{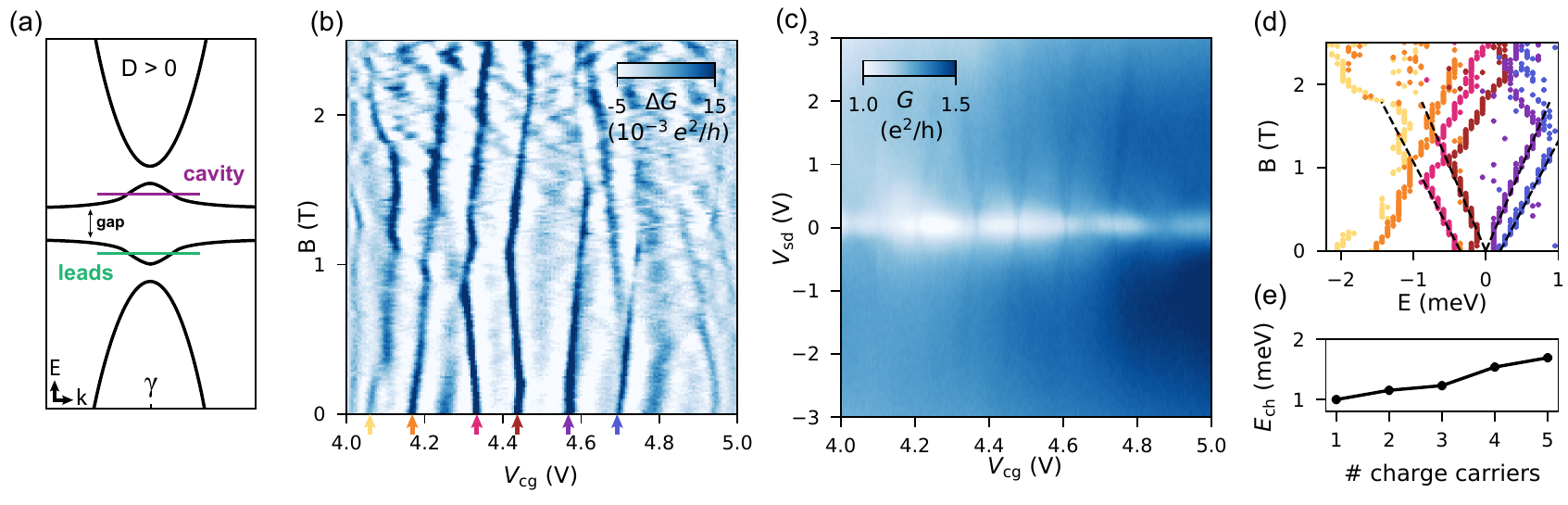}
\caption{(a) Band structure schematics. The purple line represents the Fermi energy in the cavity. The green line represents the Fermi energy in the leads. (b) Relative conductance as a function of magnetic field. We observe peaks in conductance evolving as kinked lines, characteristic of the energy levels of a quantum dot. (c) Conductance as a function of source-drain voltage and top gate voltage. We superimpose a low frequency, small amplitude AC excitation, $V{\mathrm{ac}} = \SI{100}{\mu V}$, on top of a DC signal. We observe Coulomb diamonds on top of an irregular background due to parallel conduction channels. (d) Extracted energy levels of the dot from data presented in (b). Charging energy is subtracted and the voltage axis is rescaled to energy by using the lever arm of the dot. The origin is chosen arbitrarily. Superimposed black dashed lines have a slope of $\pm g_v \pm g_s$. The best overlap is obtained for $g_v = 19$ and $g_s = 2$. (e) Charging energy of the QD as a function of the corresponding peak number, starting from the one indicated by a yellow arrow in (b).
}
\label{fig:3}
\end{figure*}

We calculate the gate--graphene capacitances using a parallel plate capacitor model (See Supplemental Material). The electron density in the leads, $n_\mathrm{l}$, and in the cavity, $n_\mathrm{c}$, and the displacement fields $D_\mathrm{l}$ and $D_\mathrm{c}$ in the same regions will be the relevant parameters for tuning the device. The three gates allow us to tune three of them independently via suitable combinations of the three possible voltage differences between pairs of gates. All measurements are performed in a $^3$He--$^4$He dilution refrigerator at a temperature of $\SI{60}{mK}$. We use a four-terminal current-biased measurement for obtaining the data shown in Figs.~\ref{fig:1},  \ref{fig:2} and \ref{fig:4}. For obtaining the data shown in Fig.~\ref{fig:3} we use a two-terminal voltage-biased measurement. Both types of measurements are performed with standard low-frequency lock-in techniques.

In Fig.~\ref{fig:1}(c), we characterize electronic transport in the device by measuring a conductance map as a function of the top and back gates. The central gate is biased together with the top gate with the intention to induce in the leads and in the cavity coinciding displacement field and density. The black arrows in Fig.~\ref{fig:1}(c) indicate the directions of increasing displacement field $D$ and density $n$. We observe low conductance regions corresponding to the gap between flat and dispersive bands (pink arrows) and to the opening of a band gap at the charge neutrality point (CNP) for a finite displacement field (blue arrows). Taking into consideration the moiré lattice and hypothesizing the presence of flat and dispersive bands, from Fig.~ \ref{fig:1}(c) we deduce the band structure at zero displacement field that we schematically represent in \ref{fig:1}(d). The main features of this schematics are confirmed in a continuum model calculation (see Supplemental Material). 


We introduce potential barriers in the system by fixing $D_{\mathrm{l}}$ and $n_{\mathrm{l}}$ and allowing the central gate to be tuned independently. The resulting conductance modulation $\Delta G$ (see Supplemental Material for a description of the background removal procedure. We refer to $\Delta G$ as the relative conductance from now on) in such a configuration is shown in the top panel of Fig.~\ref{fig:1}(e), where $D_{\mathrm{l}}= \SI{0}{V/nm}$ and $n_{\mathrm{l}} = \SI{-1e12}{cm^{-2}}$, corresponding to $V_{\mathrm{bg}}= \SI{-1.52}{V}$ and $V_{\mathrm{tg}}= \SI{-0.8}{V}$ (star in Fig.~ \ref{fig:1}(c)) and to region II in Fig.~\ref{fig:1}(d). We observe clear oscillations when the density in the cavity $n_{\mathrm{c}}$ is tuned into the dispersive bands [regions I and IV in Fig.~\ref{fig:1}(d)], independent of the electron- or hole-like dispersion of the charge carriers. To investigate their origin, we study the dependence on perpendicular magnetic field $B$. Figure~\ref{fig:1}(e) shows $\Delta G$ in the plane of center gate voltage and magnetic field. The background subtraction procedure gives rise to field-independent artifacts between the four labeled regions, but it increases the visibility of the oscillations. We observe that $B$ induces a bending of the oscillation minima and maxima and that they disappear above the orange dashed line.

This line represents the magnetic field at which the classical cyclotron radius of the charge carriers equals the lithographic length of the cavity. Fabry--Perot oscillations are expected to fade away above this threshold~\cite{Young2009, Rickhaus2015}. From this observation we conclude that forming two dispersive band/flat band interfaces in series creates the Fabry--Perot cavity in our sample. Based on this interpretation, we characterize the barriers by calculating their transparency $\mathcal{T}$ from the amplitude of the oscillations assuming an ideal Fabry--Perot cavity. We obtain $\mathcal{T}=0.33 \pm 0.02$ (see Supplemental Material). We speculate that the observation of Fabry--Perot oscillations in two dimensions could be a consequence of an anisotropic Fermi contour \cite{Gold2021} since that could provide angle selectivity \cite{Katsnelson2006,Varlet2014} . The presence of Fabry--Perot oscillations is also consistent with other known 2D interferometers which have different energy dispersion between the leads and the cavity~\cite{Karalic2020}. However, in contrast to previous observations in graphene, Fabry-Perot interferences are observed here in the absence of p-n junctions. 

To understand the effect of variations of $D_{\mathrm{c}}$ and $n_{\mathrm{c}}$ on the Fabry--Perot oscillations, we measure the conductance map shown in Fig.~\ref{fig:2}(a). The density in the leads is tuned to $n_{\mathrm{l}} = \SI{-0.8e12}{cm^{-2}}$, corresponding to the Fermi energy tuned into region II of Fig.~\ref{fig:1}(d). The value of $n_{\mathrm{l}}$ is chosen to increase the visibility of the conductance oscillations as a function of $n_{\mathrm{c}}$ in regions I and IV. We observe that they change as $D_{\mathrm{c}}$ varies, giving rise to a beating pattern. This suggests the presence of two overlapping but independent Fabry--Perot oscillations that are reminiscent of two energetically overlapping bands that vary as a function of $D_{\mathrm{c}}$~\cite{Rickhaus2019}.

In order to investigate this hypothesis, we calculate the band structure of TDBG at $1.07^{\circ}$ with a continuum model~\cite{Bistritzer12233} (see Supplemental Material for details about the model). In Fig.~\ref{fig:2}(b) Fermi surfaces for $D_{\mathrm{c}} = \SI{0}{V/nm}$ are presented for three values of the density $n_{\mathrm{c}}$. We see that at $n_{\mathrm{c}} = \SI{-3.2e12}{cm^{-2}}$ only one band is occupied (left panel), while for lower $n_{\mathrm{c}}$ a second valence band appears (central and right panels). We calculate the Fabry--Perot oscillations originating from these bands for $D_{\mathrm{c}} = \SI{0}{V/nm}$ and $D_{\mathrm{c}} = \SI{0.17}{V/nm}$  (for details see Supplemental Material). Comparing the results shown in Fig.~\ref{fig:2}(c)  in the bottom row to the experimental data in the top row, we observe that the beating pattern is qualitatively reproduced and that the order of magnitude of the periodicity matches. The correspondence is not perfect due to the simplifications of the model, such as approximating the Fermi contours of Fig.~\ref{fig:2}(b) to be circular. Nevertheless, the appearance of a beating pattern with comparable periodicity and its tunability with displacement field supports the hypothesis that the beating stems from two different bands. Consistent with this is a low mean interband scattering time, which we estimate to be larger than \SI {5}{ps} (See Supplemental Material).

We now investigate a Fabry--Perot cavity in which the Fermi energy of both the leads and the cavity are tuned into flat bands [c.f. Fig.~\ref{fig:1}(d)]. When the cavity is tuned into region III, the Fermi energy inside the cavity is in the flat band above the CNP. Because the leads are in the flat band below the CNP, with $n_{\mathrm{l}} = \SI{-0.8e12}{cm^{-2}}$ [region II, grey dashed line in Fig.~\ref{fig:1}(c)], a p-n-p cavity is formed. In this configuration, sets of weak oscillations are observed [region III, Fig.~\ref{fig:2}(a)]. Interestingly, when shifting $n_{\mathrm{l}}$ to a more negative value, namely $n_{\mathrm{l}} = \SI{-1.0e12}{cm^{-2}}$ [region II, turquoise dashed line in Fig.~\ref{fig:1}(c)], more pronounced oscillations appear [Figs.~\ref{fig:2}(d,e)]. We interpret them as Fabry--Perot oscillations originating from the formation of semitransparent barriers between regions with different charge carrier polarities. This is consistent with the absence of oscillations in region II, where the carriers both inside and outside the cavity are in the flat band below the CNP. We calculated the transparency of the barriers to be $\mathcal{T}=0.99 \pm 0.01$ (see Supplemental Material). The reason why the oscillations are clearly present for $n_{\mathrm{l}} = \SI{-1.0e12}{cm^{-2}}$ and fade away for $n_{\mathrm{l}} = \SI{-0.8e12}{cm^{-2}}$ is not clear. It is conceivable that the transparency of the barrier depends on $n_{\mathrm{l}}$. Additional possible explanations are a larger dwell-time to interband-scattering-time ratio, or a combination between a small Fermi surface in the cavity and the parallel momentum conservation rule across the interfaces~\cite{Campos2012}.

The presence of Fabry--Perot interference patterns in a cavity formed between flat bands is surprising because the increased effective mass implies a cavity dwell time an order of magnitude longer compared to the one for dispersive bands. (see Supplemental Material). A larger dwell time could increase the probability of scattering events, nevertheless the coherence of the state is maintained for at least \SI{400}{nm}, the lithographic width of the cavity. Above the line indicated by the black arrow in Fig.~\ref{fig:2}(e), the Fabry--Perot oscillations present two overlapping patterns with different slopes (white lines), while below the line only one vertical Fabry--Perot oscillation (brown line) is present. Furthermore, we observe horizontal oscillations (orange lines). We hypothesize that they originate from intersubband scattering, which is beyond the scope of this work.

To gain insight into the different Fabry--Perot patterns we compare the simulated Fermi contours at two different densities in the flat band above the CNP [Fig. \ref{fig:2}(f)]. We observe that at higher densities a Fermi contour appears around the $\gamma$ point, the center of the Brillouin zone, indicating that as the density increases the topology of the band structure changes. Therefore we interpret the line indicated by the arrow in Fig.~\ref{fig:2}(e) as a Lifshitz transition.

\section{Quantum dots}
When the leads and cavity are tuned into opposite polarity flat bands, there exists a highly resistive regime at finite displacement field $D_{\mathrm{l}}$ in the leads [Inset of Fig. \ref{fig:2}(d)]. No Fabry--Perot oscillations are present in this regime. The inset of Fig.~\ref{fig:2}(d) shows two conductance cuts at $D_{\mathrm{l}} = \SI{0.15}{V/nm}$ (grey, solid) and $D_{\mathrm{l}} = \SI{-0.1}{V/nm}$ (grey, dashed). Both show a drop in conductance when the cavity is tuned into the flat conduction band and the leads into the flat valence band. The drop along the first cut is $~10\%$ while for the second it is $~70\%$. Band structure calculations (see Supplemental Material) show that at finite displacement field, a band gap between flat bands of opposite polarity is opened, as depicted in Fig.~\ref{fig:3}(a). The Fermi energy crosses such gap at the lead/cavity interfaces, which explains the increase in resistance.

In this negative displacement field regime we observe sharp conductance resonances which we probe as a function of magnetic field, as shown in Fig.~\ref{fig:3}(b). The corresponding line cut in a $(n, D)$ map would be of the same slope as the grey lines in Fig.~\ref{fig:2}(d). The fact that the peaks do not fade away above $B \sim 0.5 \mathrm{T}$, in contrast to the oscillations shown in Fig.~\ref{fig:1}(e), rules out that they are Fabry--Perot oscillations. Since the lines show kinks but no progressive bending, we also discard snake states~\cite{Rickhaus2015}. The resonances present a linear dispersion with magnetic field, a clear signature of charge localization with corresponding energy levels that shift by the Zeeman or valley Zeeman effect~\cite{Castro_Neto2009}. Comparing these data to previous QD experiments in graphene, we believe that such peaks correspond to tunnelling events through successive energy levels of a quantum dot~\cite{Eich2018}. At zero magnetic field, the separation of peaks corresponds to the addition energy, which is the total energy needed by an electron to tunnel into the next energy level of the dot. In standard quantum dot models, the addition energy is the sum of the charging energy and the single-particle level spacing in the dot. The charging energy is the reason for which the resonances do not touch, when the magnetic field is changed~\cite{Kouwenhoven1997}.

To obtain energy resolution, we perform source-drain voltage bias measurements in the same gate voltage range, shown in Fig.~\ref{fig:3}(c). We observe, on top of a significant background conductance, Coulomb-blockade diamonds, in agreement with the strong localization hypothesis. The background conductance is expected, since we do not have a channel geometry in our device to confine all electron transport through a particular, well-defined quantum dot. From this diamond measurement we extract the lever arm of the gate (see Supplemental Material) and convert the gate voltage axis into energy [Fig.~\ref{fig:3}(d)]. We extract the gate voltage positions of the resonances at each magnetic field from Fig.~\ref{fig:3}(b) and shift them horizontally until they touch (see Supplemental Material). To a good approximation, this corresponds to subtracting the magnetic-field-independent charging energy from the resonance separation. We are then left with an approximate single-particle energy level spectrum of the dot as a function of magnetic field. From the evolution of these energy levels with magnetic field (using $E = g \mu_\mathrm{B} B$) we estimate a valley $g$-factor of 19 and a spin $g$-factor of 2, close to values obtained earlier in bilayer graphene quantum dots~\cite{Kurzmann2019}.

\begin{figure*}[t]
\centering
\includegraphics[width=1\textwidth]{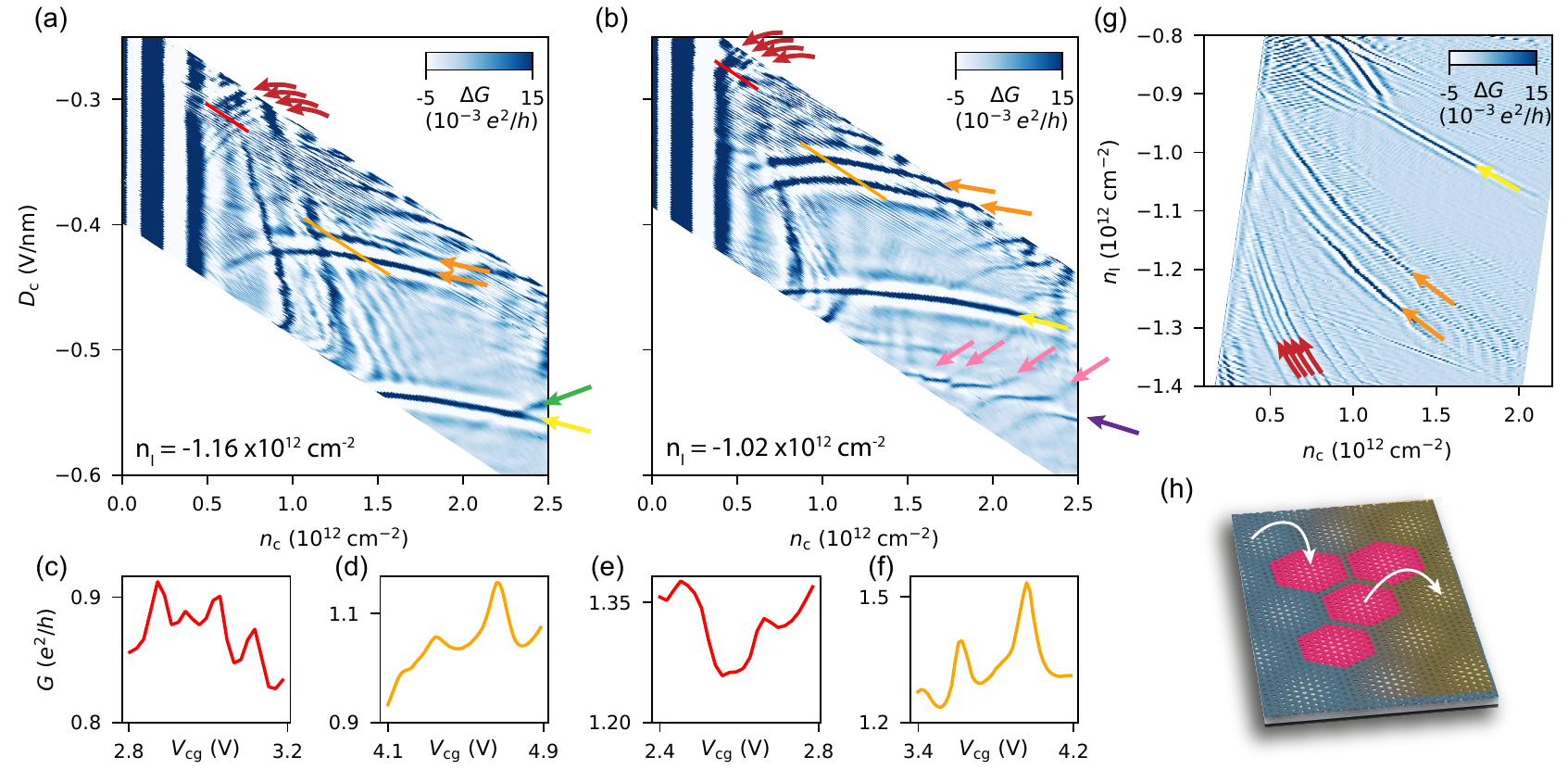}
\caption{(a) Relative conductance as a function of displacement field and electron density in the cavity. The density in the leads is fixed at $n_{\mathrm{l}} = \SI{-1.16e12}{cm^{-2}}$. Different sets of Coulomb conductance oscillations (red, orange, yellow and green arrows) are observed. Red and orange lines indicate cuts in top gate voltage along which conductance is shown in (c) and (d). (b) Same figure as (a) with a different value for the electron density in the leads $n_{\mathrm{l}} = \SI{-1.02e12}{cm^{-2}}$. Pink and violet arrows highlight avoided crossing feature. (c)-(f) Conductance as a function of top gate voltage along cuts highlighted in (a) and (b). (g) Relative conductance as a function of electron densities in and outside the cavity. The resonances in (a) and (b) are highlighted by the same colors. (h) Schematic representation of a 'moiré atom'. An electron density gradient (blue to yellow) combined with the superlattice leads to charge islands forming at the moiré unit cells (pink hexagons).
}
\label{fig:4}
\end{figure*}

Figure \ref{fig:3}(e) shows the charging energy for each state of the QD. 
It increases as a function of voltage, suggesting holes to be the charge carriers in the dot. In addition, it shows a significant slope. Ref.~\cite{Eich2018} reports a charging energy nearly independent of the electron number in the dot from 20 electrons onwards. Therefore, we consider it likely that our dot contains between 1 and 20 holes. From the extracted values and a parallel plate disk capacitor model for the QD we obtain a crude upper bound for the diameter of the dot of $\SI{200}{nm}$ (see Supplemental Material). Thus, the dot, if circular, covers at most $\SI{0.2}{\mu m}/\SI{2}{\mu m} = 10\%$ of the interface, in agreement with the large background current in Figs.~\ref{fig:3}(a), \ref{fig:3}(b).

We now shift our focus on understanding the spatial extent and location of the dots better. We perform current biased conductance measurements in a parameter region where diamonds are observed. Figures~\ref{fig:4}(a) and \ref{fig:4}(a)(b) show conductance data $G(n_{\mathrm{c}}, D_{\mathrm{c}})$ taken at fixed lead densities of $n_{\mathrm{l}} = \SI{-1.16e12}{cm^{-2}}$ and $n_{\mathrm{l}} = \SI{-1.02e12}{cm^{-2}}$, respectively. 
Sets of prominent oscillations (red, orange and yellow arrows) arise. As shown in Figs.~\ref{fig:4}(c) to \ref{fig:4}(f), their relative height in conductance is close to 10\%, significantly higher than the 0.2\% value of the Fabry--Perot oscillation amplitude in Fig.~\ref{fig:2}(e). We see that changing the lead electron density only has a weak effect on the relative amplitude of the peaks. However, it shifts the sets of resonances in (n,D)-space significantly. We believe that each set of oscillations corresponds to a set of energy levels of an individual QD. These sets of resonances are distributed in $n_\mathrm{c}$ and $D_\mathrm{c}$ [Figs.~\ref{fig:4}(a,b)], and their relative evolution differs from one set to another. Figure~\ref{fig:4}(a) shows the crossing of two resonances (yellow and green arrows), pointing to spatially separated charged islands. On the other hand, Fig.~\ref{fig:4}(b) shows avoided-crossings between different resonances (pink and purple arrows). This pattern resembles resonances of double QDs~\cite{Eich2018_2} and points to the formation of quantum dots which are in spatial proximity.

To narrow down the spatial region in which charge localization takes place, we measure conductance maps [Fig.~\ref{fig:4}(g)] as a function of electron densities $n_{\mathrm{c}}$ and $n_{\mathrm{l}}$ inside and outside the cavity. We observe the same sets of resonances, this time in the form of straight or slightly bent lines, depending on the oscillation set. The resonances are tuned by the densities in both the cavity and the leads. This points to charges being localized at the interface between the two regions. The upper bound estimate of $\SI{200}{nm}$ for the diameter of the dots, if not elongated, is half of the lithographic width of the gate defining the cavity. This rules out the situation in which a $\SI{400}{nm}$ wide quantum dot is formed under the entire width of the gate. In addition, we observe in Figs.~\ref{fig:4}(a), \ref{fig:4}(b) a vertical CNP line at $n_\mathrm{c} = 0$, obtained from our plate capacitor model. On the contrary, the mentioned resonances are not straight vertical lines. This speaks for a different capacitance between the dots and the central gate than the one expected if the dots were fully under the gate. We observe $dn_\mathrm{l}/dn_\mathrm{c}$ slopes of approximately -0.8 (red), -0.4 (orange) and -0.2 (yellow). This can be interpreted as the QD responsible for the red set of oscillations being further away from the leads than the one responsible for the orange resonances, in turn being further away than the one responsible for the yellow resonances. When plotting the relative conductance as a function of gate voltage, slopes can be interpreted in terms of relative lever arms of the different gates to the dots (see Supplemental Material). They, in turn, depend mainly on two factors: the density gradient across the interface, determining the potential landscape felt by the dot, and the exact position of the dot with respect to the interface.

The Fermi energy in the sample passes through a gap at the interface between cavity and leads. We estimate the length of the high electric potential gradient region to be about \SI{30}{nm}, which is the thickness of the top hBN layer. This is to be compared with the size of the moiré unit cell, which, at our twist angle, is $L_{\mathrm{m}} = \SI{13.2}{nm}$. An electron going through the interface therefore feels the electrostatic potential modulation by only a few moiré unit cells. Recent scanning tunnelling spectroscopy (STS) studies in magic-angle twisted bilayer graphene showed an enhancement of the effect of local potential modulations on the charge carriers in flat bands, leading to quantum dot formation~\cite{Tilak2021}. This points towards the flatness of the bands, a property common to many moiré materials, as being at the origin of the localization mechanism.

We estimate the spatial extent of the dot in terms of moiré unit cells. Such moiré unit cells, accumulating carriers, can be regarded as a 'moiré atom' where electrons tunnel in and out, as depicted in Fig.~\ref{fig:4}(h). These atoms can host, however, a limited number of charge carriers. We know that the densities at which we observe the Coulomb resonances in our data are below $n = \SI{3e12}{cm^{-2}}$, the density at which the flat bands are completely filled. At that density, assuming spin and valley degeneracy, each moiré atom hosts 4 charge carriers~\cite{Cao2018_1}. We count at least 8 resonances in Fig.~\ref{fig:3}(a), meaning that at least two moiré atoms form the corresponding charged island. This, combined with the size of the moiré unit cells and the upper bound of the size of the dots, lets us estimate that the Coulomb resonances stem from islands formed out of two to ten moiré atoms.

\section{conclusion}
In this work, we characterized the different electronic confinement regimes in a TDBG cavity. They stem from the highly tunable band structure resulting from the moiré lattice of the material. Weak confinement was induced through the formation of two novel types of Fabry--Pérot cavities. In one of them, the barriers are dispersive/flat band interfaces, and interference is observed also in the absence of a p-n junction. In the second, junctions arise between spatial regions with flat bands, demonstrating preservation of coherence across the cavity despite the high effective masses. Furthermore, strong charge localization was observed along the cavity interfaces in the flat band regime. Localized islands extending across several moiré unit cells show a behavior reminiscent of Bernal-stacked bilayer graphene quantum dots. Such interfacial structures can lead to unwanted charge trapping in gate-defined single crystal devices, hampering their quality. On the other hand, such localization, if well controlled, can open new possibilities to engineer quantum devices in moiré materials. A better understanding and harnessing of electron localization at electrostatically defined interfaces seems to be a crucial step in the further development of moiré-based devices.

The data used in this Letter will be made available through the ETH Research Collection.

\begin{acknowledgements}
We acknowledge the support from Peter Maerki, Thomas Baehler and the staff of the ETH FIRST cleanroom facility.
We acknowledge support from the Graphene Flagship, the Swiss National Science Foundation via NCCR Quantum Science and from the European Union's Horizon 2020 research and innovation programme under grant agreement No 862660/QUANTUM E LEAPS.
E.P acknowledges support of a fellowship from ”la Caixa” Foundation (ID 100010434) under fellowship code LCF/BQ/EU19/11710062.
K.W. and T.T. acknowledge support from the Elemental Strategy Initiative conducted by the MEXT, Japan, Grant Number JPMXP0112101001,  JSPS KAKENHI Grant Number JP20H00354 and the CREST(JPMJCR15F3), JST.
E.P. and G.Z. contributed equally to this work.
\end{acknowledgements}

\end{document}


\title{Supplemental Material for Fabry-Pérot cavities and quantum dot formation at gate-defined interfaces in twisted double bilayer graphene}

\author{El\'ias Portolés$^{\dagger}$}
\email{eliaspo@phys.ethz.ch}
\author{Giulia Zheng}
\email{These authors contributed equally}
\author{Folkert K. de Vries}
\affiliation{Solid State Physics Laboratory, ETH Zürich,~CH-8093~Zürich, Switzerland}
\author{Jihang Zhu}
\affiliation{Department of Physics, University of Texas at Austin, Austin, Texas 78712, USA}
\author{Petar Tomić}
\affiliation{Solid State Physics Laboratory, ETH Zürich,~CH-8093~Zürich, Switzerland}
\author{Takashi Taniguchi}
\affiliation{International Center for Materials Nanoarchitectonics,
National Institute for Materials Science,  1-1 Namiki, Tsukuba 305-0044, Japan}
\author{Kenji Watanabe}
\affiliation{Research Center for Functional Materials,
National Institute for Materials Science, 1-1 Namiki, Tsukuba 305-0044, Japan}
\author{Allan H. MacDonald}
\affiliation{Department of Physics, University of Texas at Austin, Austin, Texas 78712, USA}
\author{Klaus Ensslin}
\author{Thomas Ihn}
\author{Peter Rickhaus}
\affiliation{Solid State Physics Laboratory, ETH Zürich,~CH-8093~Zürich, Switzerland}

\maketitle

\section*{Determining the twist angle}

\begin{figure*}[h!]
\centering
\includegraphics[width=1\textwidth]{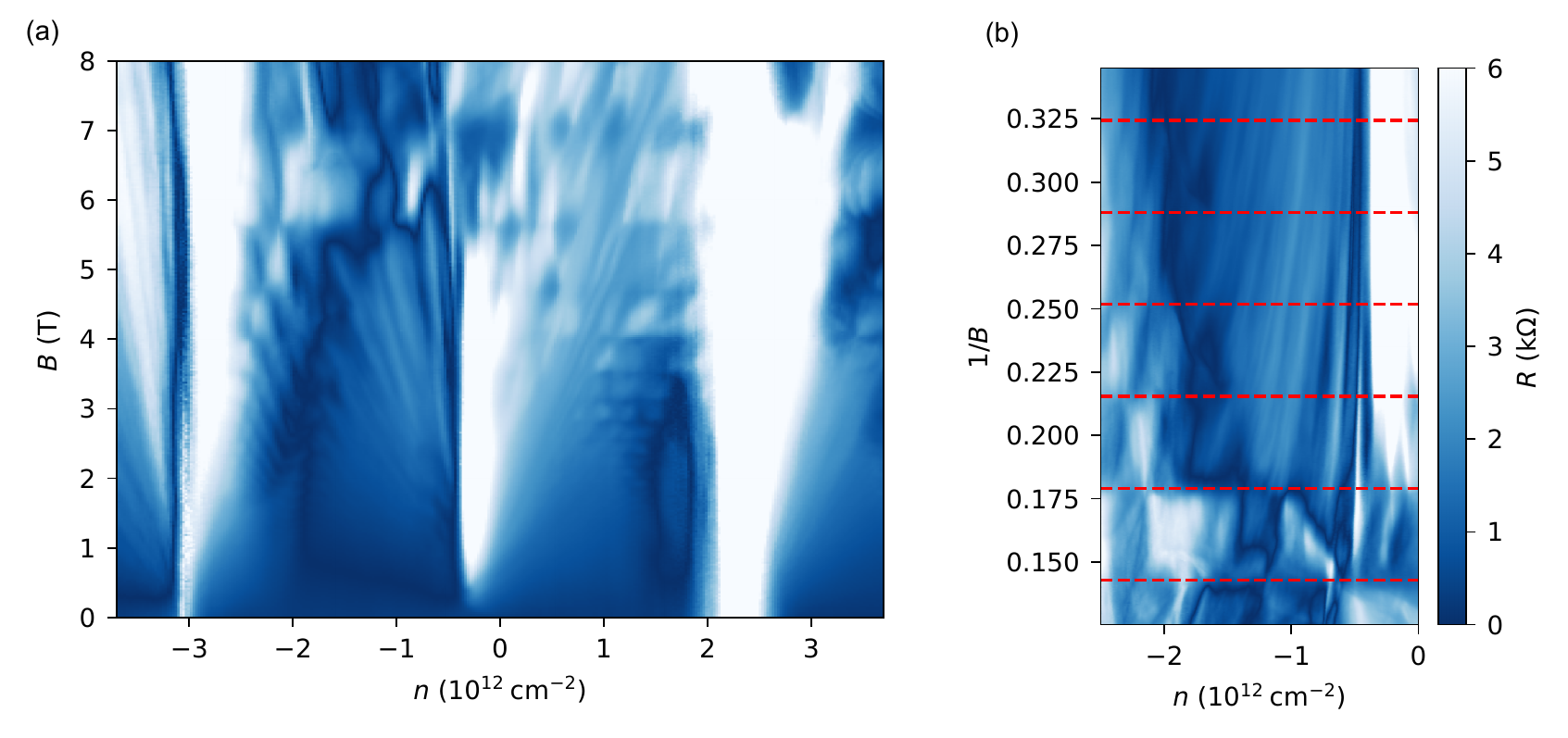}
\caption{(a) Resistance across the sample as a function of electronic density and magnetic field. Brown-Zak oscillations are visible as horizontal lines. (b) Zoom in the data from (a) and plotted as a function of $1/B$. The dashed red lines highlight the ABZ oscillations. Both figures share the same colorbar.}
\label{fig:S1}
\end{figure*}

The twist angle $\theta$ of the device is related to the lattice constant of graphene, $a$ and the moiré length $L_{\mathrm{m}}$ through the following equation~\cite{Cao2018_1}:

\begin{equation}
    \theta = 2 \arcsin\left(\frac{a}{2L_{\mathrm{m}}}\right)
\end{equation}

where $L_{\mathrm{m}} = \sqrt{2 \mathcal{A} / \sqrt{3}}$. $\mathcal{A}$ is the area of the moiré unit cell, which we estimate from Azbel-Brown-Zak (ABZ) oscillations~\cite{Azbel1961, Brown1964, Zak1964} measured in our device. Figure \ref{fig:S1}(a) shows resistance as a function of magnetic field and electronic density. We observe resistance oscillations which are independent of density. These are ABZ oscillations. Figure \ref{fig:S1}(b) shows a zoom of the data plotted as a function of $1/B$. When the number of moiré unit cells threaded by one magnetic flux quantum equals an integer, the ABZ oscillations present a maximum in conductance. From this, we compute:

\begin{equation}
    \Delta \left(\Phi_0 / \Phi \right) = \frac{\Phi_0}{\mathcal{A}} \Delta\left(\frac{1}{B} \right) = 1 \implies \mathcal{A} = \Phi_0\Delta\left(\frac{1}{B} \right)
\end{equation}

We measure $\Delta(1/B) \sim 0.036$ $\mathrm{T^{-1}}$ and obtain $\mathcal{A} = \SI{150}{nm^2}$, from which we compute a twist angle of $1.07^{\circ}$.

\clearpage
\newpage

\section*{Density and displacement field calculation}

We determine the gate capacitances to the graphene from a parallel plate capacitor model, yielding $C_{\mathrm{cg}} = \varepsilon_0 \varepsilon_{\mathrm{hBN}}/d_{\mathrm{top}}$, $C_{\mathrm{tg}} = \varepsilon_0 \varepsilon_{\mathrm{hBN}}/(d_{\mathrm{top}} + d_{\mathrm{Al_2O_3}})$ and $C_{\mathrm{bg}} = \varepsilon_0 \varepsilon_{\mathrm{hBN}}/d_{\mathrm{bottom}}$ for the center, top and bottom gates respectively. $d_{\mathrm{top}}$, $d_{\mathrm{bottom}}$ and $d_{\mathrm{Al_2O_3}}$ are the thicknesses of the top hBN flake, bottom hBN flake and $\mathrm{Al_2O_3}$ layer respectively and $\varepsilon_0$ and $\varepsilon_{\mathrm{hBN}}$ represent the vacuum permitivity and the relative permitivity of hBN. We obtain the electronic densities in the central and lead regions $n_{\mathrm{c/l}} = (C_{\mathrm{cg/tg}} V_{\mathrm{cg/tg}} + C_{\mathrm{bg}} V_{\mathrm{bg}})/e$, and the displacement fields, $D_{\mathrm{c/l}} = (C_{\mathrm{cg/tg}} V_{\mathrm{cg/tg}} - C_{\mathrm{bg}} V_{\mathrm{bg}})/2\varepsilon_0$~\cite{Rickhaus2019}.

\clearpage
\newpage

\section*{Bandstructure calculation of TDBG}
The low-energy continuum model of TDBG is constructed by extending the continuum model of twisted bilayer graphene (TBG)\cite{Bistritzer12233}. In TDBG, the middle two graphene layers are twisted by a small angle $\theta$ with respect to each other and follow the same Hamiltonian model as TBG. The topmost two graphene layers and the bottom most two graphene layers are described by the minimum model of Bernal-stacked bilayer graphene, i.e. only the interlayer tunneling between dimer sites is included. The Hamiltonian of TDBG in momentum space, expanded in the Bloch states
\begin{equation}
\psi^{(l)}_{\mathbf{k} \alpha} (\mathbf{r}) = \frac{1}{\sqrt{N}} \sum\limits_{\mathbf{R}} e^{i\mathbf{k} \cdot (\mathbf{R}+\boldsymbol{\tau}_\alpha)} \phi(\mathbf{r}-\mathbf{R}-\boldsymbol{\tau}_\alpha)
\end{equation}
for each layer indexed by $l$, is
\begin{equation}
\label{Eq_Hamil}
H(\mathbf{k}) = 
\begin{pmatrix}
h_{\theta/2}(\mathbf{k}+\mathbf{g}-\mathbf{K}_t) & t_1 & 0 & 0 \\
t_1 & h_{\theta/2}(\mathbf{k}+\mathbf{g}-\mathbf{K}_t) & T_{\mathbf{g}\mathbf{g}'} & 0 \\
0 & T^\dagger_{\mathbf{g}\mathbf{g}'} & h_{-\theta/2}(\mathbf{k}+\mathbf{g}'-\mathbf{K}_b) & t_1 \\
0 & 0 & t_1 & h_{-\theta/2}(\mathbf{k}+\mathbf{g}'-\mathbf{K}_b) &
\end{pmatrix}
\end{equation}
where $\mathbf{g}$ and $\mathbf{g}'$ are moir\'e reciprocal lattice vectors. We cutoff $\mathbf{g}$ till the fourth shell, i.e. $|g_{\text{max}}| = 4g$ and $g$ is the length of the primitive reciprocal lattice vector.
$\mathbf{K}_t$ and $\mathbf{K}_b$ are Dirac points of the top and bottom bilayer graphene after twisting. Near graphene's Brillouin zone corner $\mathbf{K}_+=(1,0)4\pi/3a$, where $a=0.246$ nm is the lattice constant of graphene,
\begin{equation}
\mathbf{K}_t = \mathcal{R}_{\theta/2}\mathbf{K}_+, \qquad
\mathbf{K}_b = \mathcal{R}_{-\theta/2}\mathbf{K}_+,
\end{equation}
$\mathcal{R}_\theta$ is the rotation matrix. In Eq.~(\ref{Eq_Hamil}), $h_\theta(\mathbf{q})$ is the Dirac Hamiltonian rotated by $\theta$:
\begin{equation}
h_\theta(\mathbf{q}) = \hbar v_F q
\begin{pmatrix}
\Delta & e^{-i(\theta_\mathbf{q} - \theta)} \\
e^{i(\theta_\mathbf{q} - \theta)} & \Delta
\end{pmatrix}
\end{equation}
$\theta_\mathbf{q}$ is the angle of crystal momentum $\mathbf{q}$ measured from the Dirac point. $\Delta$ is the on-site energy. The interlayer tunneling matrix $T_{\mathbf{g}\mathbf{g}'}$ between the middle two graphene layers is
\begin{equation}
T_{\mathbf{g}\mathbf{g}'} = w_0 \sum\limits_{j=1}^3 T_j \delta_{\mathbf{g}-\mathbf{g}',\tilde{\mathbf{g}}_j}
\end{equation}
$\tilde{\mathbf{g}}_j$ $(j=1,2,3)$ are three interlayer hopping momentum boosts
\begin{equation}
\begin{split}
&\tilde{\mathbf{g}}_1= (0,0) \\
&\tilde{\mathbf{g}}_2= (\frac{1}{2},\frac{\sqrt{3}}{2}) \frac{4\pi}{\sqrt{3}a_M} \\
&\tilde{\mathbf{g}}_3= (-\frac{1}{2},\frac{\sqrt{3}}{2}) \frac{4\pi}{\sqrt{3}a_M} \\
\end{split}
\end{equation}
$a_M \approx a/\theta$ is the moir\'e lattice constant. Three corresponding interlayer tunneling matrices are
\begin{equation}
\begin{split}
&T_1 = 
\begin{pmatrix}
u_0 & u_1 \\
u_1 & u_0
\end{pmatrix} \\
&T_2 =
\begin{pmatrix}
u_0e^{i\phi} & u_1 \\
u_1e^{-i\phi} & u_0e^{i\phi}
\end{pmatrix} \\
&T_3 = 
\begin{pmatrix}
u_0e^{-i\phi} & u_1 \\
u_1e^{i\phi} & u_0e^{-i\phi}
\end{pmatrix} 
\end{split}
\end{equation}
where $\phi = 2\pi/3$. As a consequence of strain and corrugation effects, $u_0 < u_1$.

In the calculations in this paper, we have adopted $t_1=330$ meV, $\hbar v_F = 658$ meV$\cdot$nm which corresponds to $v_F=10^6$ m/s, interlayer tunneling strength $w_0 = 110$ meV, $u_0 = 0.8$ and $u_1=1$.

\begin{figure*}[h!]
\centering
\includegraphics[width=0.8\textwidth]{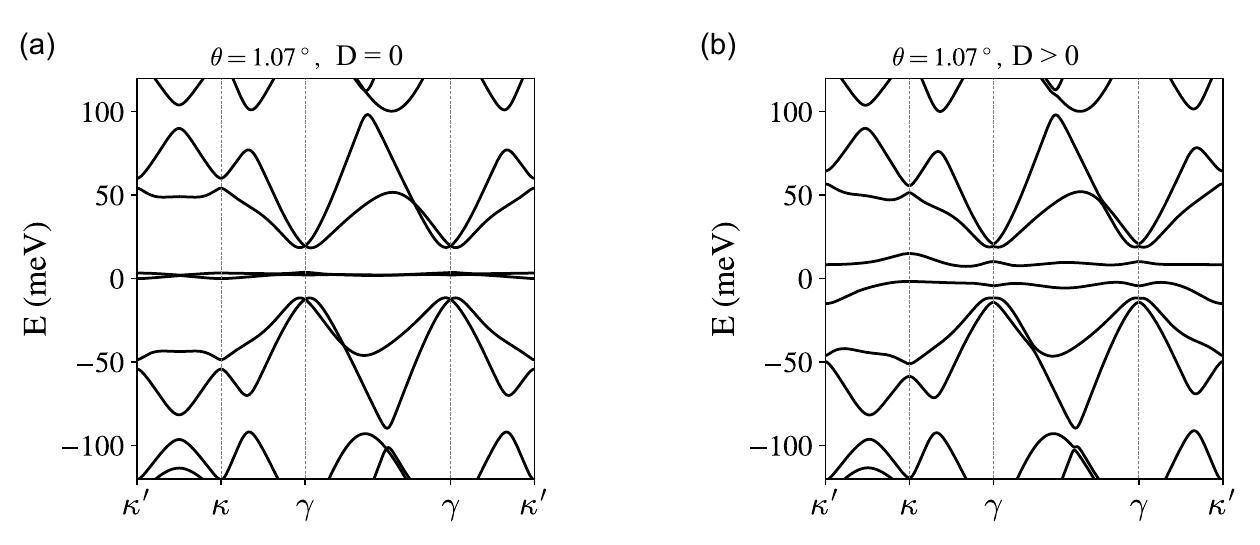}
\caption{Band structure of tDBG at $1.07^\circ$ at zero (a) and finite (b) displacement field.}
\label{fig:DOS}
\end{figure*}

\clearpage
\newpage

\section*{Background removal procedure}

\begin{figure*}[h!]
\centering
\includegraphics[width=1\textwidth]{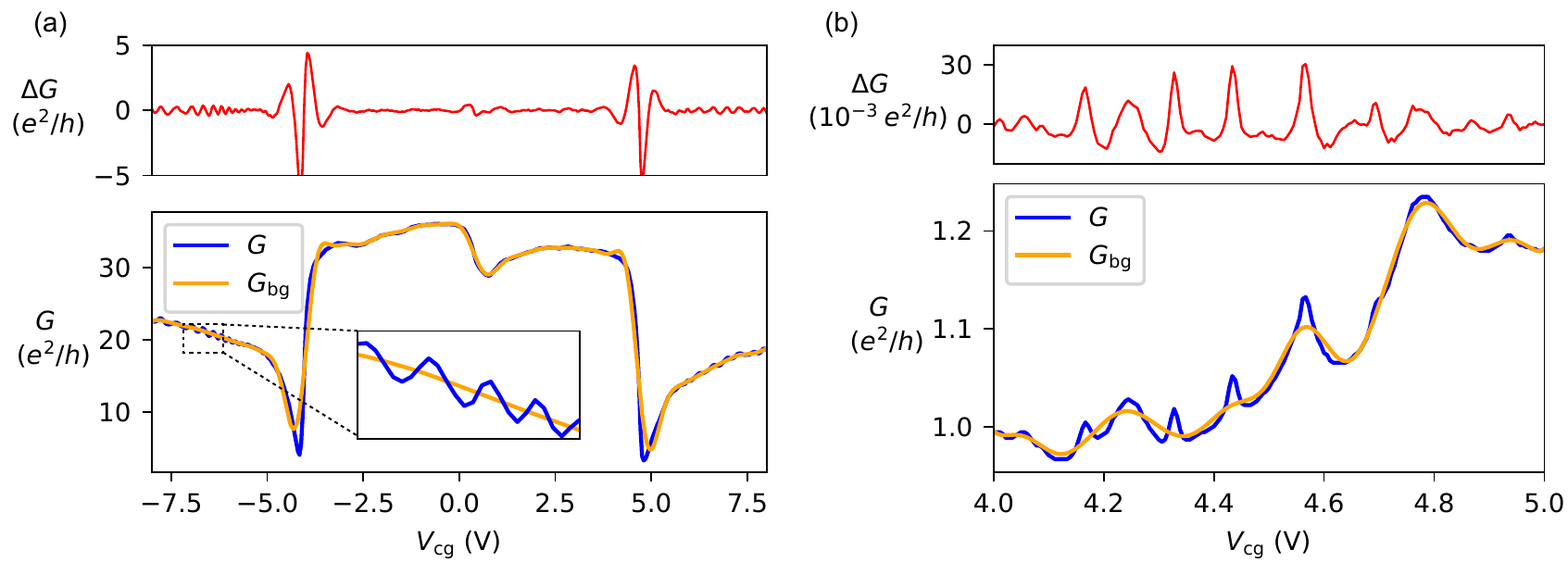}
\caption{(a) Conductance cut from Fig.~1(e) at $B = 0$. Top panel: relative conductance obtained after the background extraction procedure. Bottom panel: In blue the raw conductance data and in orange the computed background. (b) Conductance cut from Fig.~3(b) at $B = 0$. Same panel distribution and legend as (b).}
\label{fig:S7}
\end{figure*}

In order to more clearly observe peaks or oscillations in our conductance data we extract the background from it. To do so, we use a Savitzky-Golay filter~\cite{Savitzky1964}. It consists of a moving frame, low order, polynomial fit. We use a frame of 31 points and a polynomial of degree 5 for the background removal. We then iterate the procedure 4 times for data shown in Fig.~1 and Fig.~2 and 8 times for that in Fig.~3 and Fig.~4. Two examples of such extraction procedure are shown in figure \ref{fig:S7}. From applying this filtering to the conductance data $G$ we obtain a smooth background $G_{\mathrm{bg}}$. We subtract this to the original data to obtain what we refer to in the main text as relative conductance $\Delta G = G - G_{\mathrm{bg}}$.

\clearpage
\newpage

\section*{Simulation of Fabry--Pérot oscillations}
The interference pattern of a Fabry--Pérot interferometer is given by the succession of Lorentzian curves with peaks distanced $2 k W$ with respect to each other, where $k$ is the wave vector of the interfering particles and $W$ the width of the cavity~\cite{Shytov2008}. We approximate in our simulations the series of Lorentzians expected from a Fabry--Pérot interferometer with a sinusoidal function respecting the same periodicity ($\sin{2 k W}$). This approximation is justified for two reasons. First, the experimental data is compatible with the sinusoidal shape of the oscillations. Second, we are interested in comparing the period of the oscillations and not the amplitude. In order to compare it to the collected data we calculate an approximated relation between $n$ and $k$ by fitting an isotropic and parabolic curve  $n = a k^2 + c$ to the band structure, where $a$ and $b$ are fitting parameters. Additionally, we multiply this formula by a scaling factor. This compensates a difference between the calculated bandstructure and the data in terms of the density range from the charge neutrality point to the full filling gap. We perform the fitting for the two bands ($v1$ and $v2$) that are present in the range of densities that interest us. By inverting the formula we obtain $k_{v1/v2}$. Assuming that there is low intersubband scattering we obtain the Fabry--Pérot interference pattern by adding the contributions of each band: $\sin{2 k_{v1} W} + \sin{2 k_{v2} W}$.

\clearpage
\newpage

\section*{Scattering time and cavity dwell time}
For the coherence of the charges inside a cavity to be maintained, the inelastic scattering time $t_s$ must be larger than the cavity dwell time $t_d = 2 \cdot W / v_F$, where $W$ is the width of the cavity and  $v_F =\frac{1}{ \hbar} \frac{ \partial {E_F}} {\partial {k}} = \frac{\hbar k_F}{m*} $ the Fermi velocity. A high effective mass indicates a longer dwell time, therefore the flatter the band the longer needs $t_s$ to be in order for the coherence to be maintained. We approximate the Fermi velocity to be $v_F=\frac{\Delta E}{\hbar \Delta k}$. For the dispersive band we read the values of $\Delta E$ and $\Delta k$ from the band structure at zero displacement field around the $\gamma$ point (Fig.~\ref{fig:DOS}), calculate $t_d$ and obtain $t_d = \SI{5}{ps}$.  For the flat band we take $\Delta E$ to be the bandwidth of the flat band and obtain $t_d = \SI{40}{ps}$. This is a very rough approximation but allows us to estimate the order of magnitude of $t_d$, serving as the lower bound of $t_s$.

\clearpage
\newpage

\section*{Calculation of the transmission amplitude}
\begin{figure*}[h!]
\centering
\includegraphics[width=0.7\textwidth]{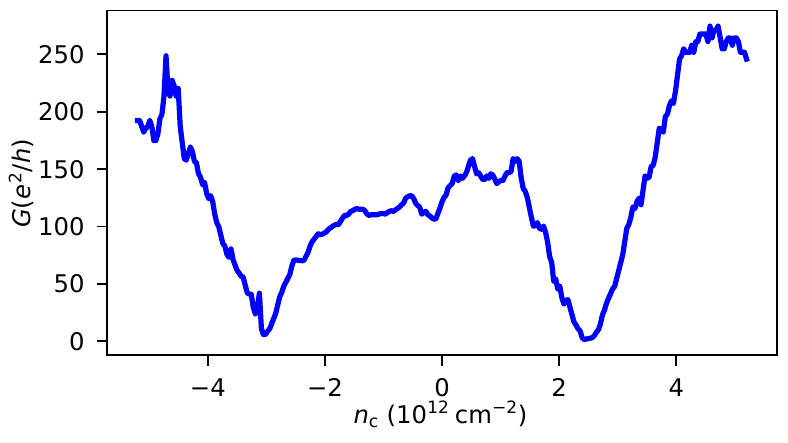}
\caption{Conductance G as a function of density when no barrier is formed between the leads and the cavity}
\label{fig:S6}
\end{figure*}

The transmission amplitude $\mathcal{T}$ for a given density $n_c$ in a ideal Fabry--Pérot cavity is estimated from the relative amplitude of the oscillations $\Delta G=N G_0 \frac{F}{2+3F+F^2} $, where the finesse $F$ is given by $F= \frac{4(1-\mathcal{T})^2}{\mathcal{T}^4}$, $N$ is the number of channels and $G_0=2e^2/h$ is the conductance quantum~\cite{Rickhaus2013}.
Therefore, when $N$ is known, one can calculate $\mathcal{T}$ from the experimental value of $\Delta G$. $N$ is read from the absolute conductance $G = N G_0 \mathcal{T}^2$, when the transmission probability $\mathcal{T}^2$ is equal to one. This is the case when the density in the cavity is tuned to be equal to the density in the leads and no barrier is formed (Fig.~\ref{fig:S6}). The transmission for the dispersive band/flat band barrier is calculated for $n = \SI{-3.5e12}{cm^{-2}}$, at this value $N = 67$ and we obtain two values of $\mathcal{T}$, $\mathcal{T}_1=0.33$ and $\mathcal{T}_2=0.98$. Because electrons have to cross a gap between flat bands and dispersive bands we expect the transmission to be significantly lower than $1$, therefore we discard $\mathcal{T}_2$. For the p-n interface in the flat bands $\mathcal{T}$ is calculated for $n = \SI{0.76e12}{cm^{-2}}$, where $N = 141$. We obtain $\mathcal{T}_2=0.19$ and $\mathcal{T}_2=0.98$. In this configuration the barrier is gapless and we expect a transmission close to $1$, therefore we discard $\mathcal{T}_1$.
The largest source of error arises from $G$ as the background conductance is not homogeneous. To estimate the consequence on the $\mathcal{T}$ value we re-performed the calculation substracting 25 to the previously used $N$. Despite the large number, we observed a difference in the value of $\mathcal{T}$ of only $0.02$ for the dispersive band/flat band barrier and $0.01$ for the p-n interface in the flat bands. 

\clearpage
\newpage

\section*{Lever arm extraction}

\begin{figure*}[h!]
\centering
\includegraphics[width=1\textwidth]{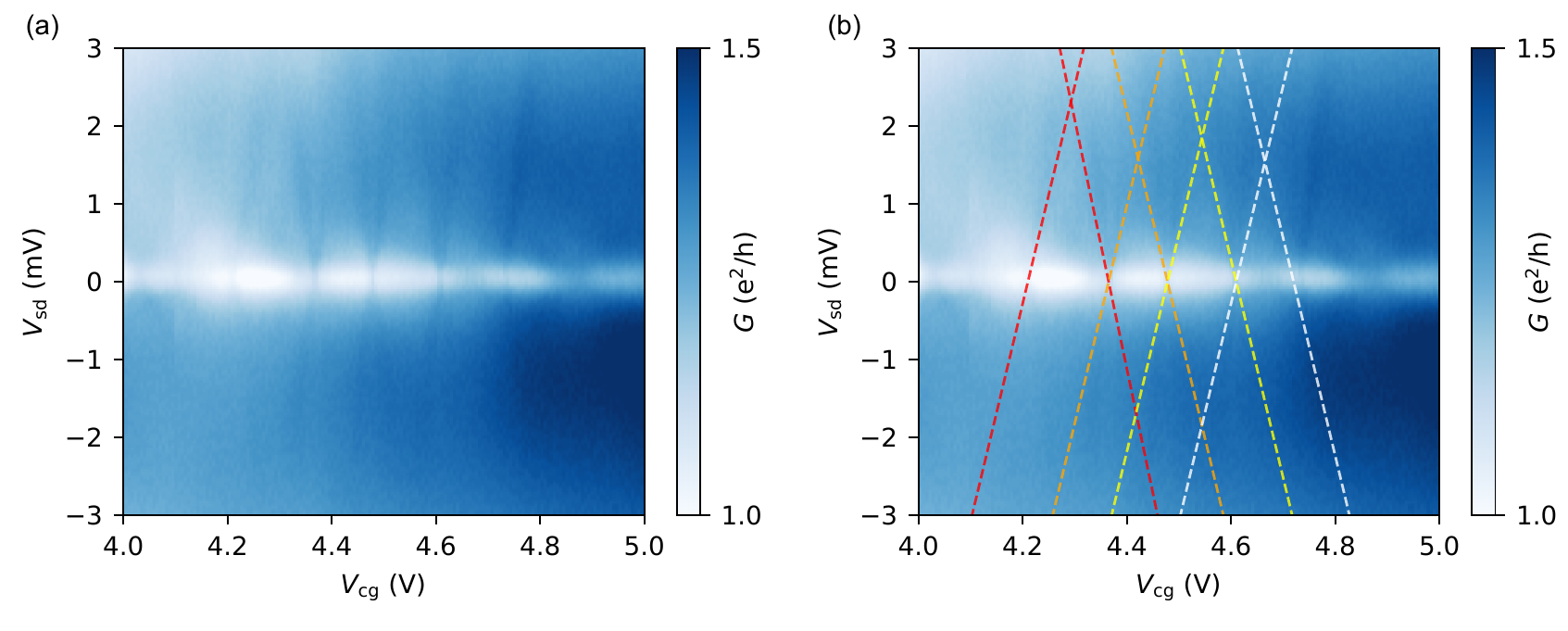}
\caption{(a) Conductance as a function of source-drain voltage and central gate voltage. We superimpose a low frequency, small amplitude AC excitation, $V{\mathrm{ac}} = 100\mu V$, on top of a DC offset. We observe Coulomb diamonds on top of an irregular background due to parallel conduction channels. This is the data presented in the main text, for comparison purposes with (b). (b) Same data with dashed lines overlapping the Coulomb diamonds serving as a guide to the eye.}
\label{fig:S2}
\end{figure*}

Each high conductance line defining a Coulomb diamond [Fig.~\ref{fig:S2}(a,b)] corresponds to either the Fermi energy at the source or the drain contacts being aligned with an energy level of the dot. At the point where two such lines of opposite slope intersect, at finite source-drain voltage, an energy level of the dot is aligned with the Fermi energy of the source (drain) while the next one is aligned with the Fermi energy of the drain (source). Therefore, the source-drain voltage at which this intersection takes place gives the energy spacing between the two consecutive energy levels, $\Delta E$, in electronvolts. At the same time, when no source-drain voltage is applied, tuning the plunger gate voltage, $V_\mathrm{cg}$ in our case, capacitively tunes the offset of all the energy levels of the dot. Therefore, by reading the voltage difference $\Delta V_\mathrm{cg}$ in plunger gate between two intersecting points at $V_{\mathrm{sd}} = 0$ one obtains the voltage needed to bridge two consecutive energy levels of the dot~\cite{Ihn2009}. The proportionality factor $\alpha = \Delta E_{\mathrm{eV}} / \Delta V_{\mathrm{cg}}$, which is half the slope of the dotted lines, is the so-called lever arm. We obtain from our measurement a lever arm $\alpha_\mathrm{cg} = \SI{14}{eV/V}$. We obtain negligible variations depending on the diamond used for computing the value, as expected~\cite{Ihn2009}.

\clearpage
\newpage

\section*{Estimating the size of the dot}
We consider a parallel plate capacitor model for estimating the size of the dots from the charging energy. The capacitance of the island is given by $C = E_\mathrm{ch}/e$, $E_\mathrm{ch}$ being the charging energy of the dot and $e$ the elementary charge~\cite{Ihn2009}. We represent the dot by a metallic circular island separated from the top and bottom gates by an insulator. In such a picture, the area of the island is given by:

\begin{equation}
    \mathcal{A} = \frac{C}{\varepsilon_0 \varepsilon_{\mathrm{hBN}} (1/d_\mathrm{top} + 1/d_\mathrm{bottom})},
\end{equation}

with $d_\mathrm{top} = \SI{30.5}{nm}$, $d_\mathrm{bottom} = \SI{54}{nm}$, from which we deduce a radius of $\SI{100}{nm}$. By considering the stack to be a circular plate capacitor we neglect the contribution to the capacitances of the parts of the back gate and central gate that are not strictly underneath or on top of the metallic disk. We also neglect the capacitance between the disk and the other gates and contacts of the device. These are not negligible, but we do not take them into account to avoid speculation about the exact capacitive coupling between the dot and the gates. Their contribution would decrease the size estimation of the dot.

\clearpage
\newpage

\section*{Relative lever arms of the gates to the dots}

\begin{figure*}[h!]
\centering
\includegraphics[width=0.5\textwidth]{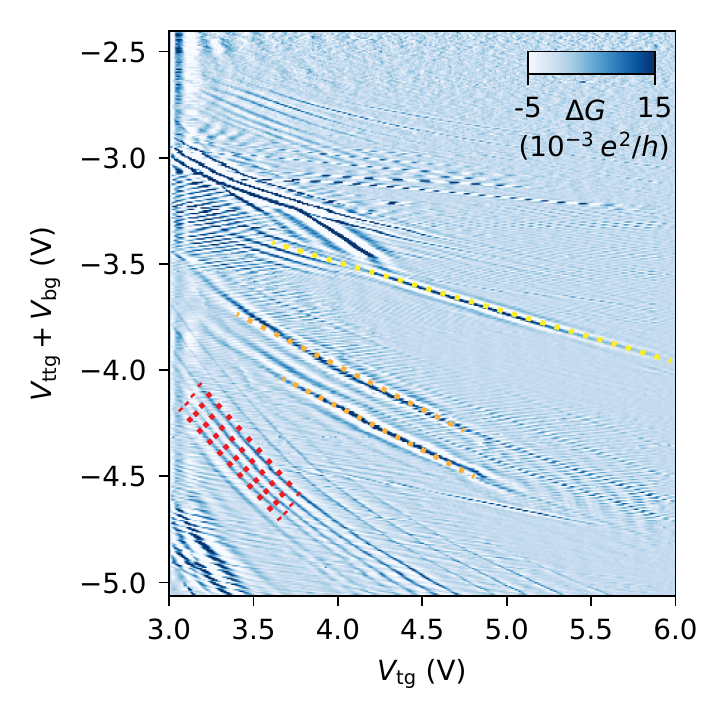}
\caption{Same data as the one presented in the main text, this time with voltages in the axis instead of densities. The slope of each resonance gives the relative lever arm of each quantum dot with respect to the central gate on one hand and the bottom and top gate on the other.}
\label{fig:S3}
\end{figure*}

The different dotted lines shown in Fig. \ref{fig:S3} give a rough estimation of the lever arm ratio between the central gate to the combined top and bottom gates for each quantum dot. Here we read, for the quantum dots referenced as yellow, orange and red in the main text lever arm ratios $\alpha_\mathrm{c}/\alpha_\mathrm{t+b}$ of 4, 2 and 1 respectively.

\clearpage
\newpage

\section*{Extended range coulomb diamonds}

\begin{figure*}[h!]
\centering
\includegraphics[width=1\textwidth]{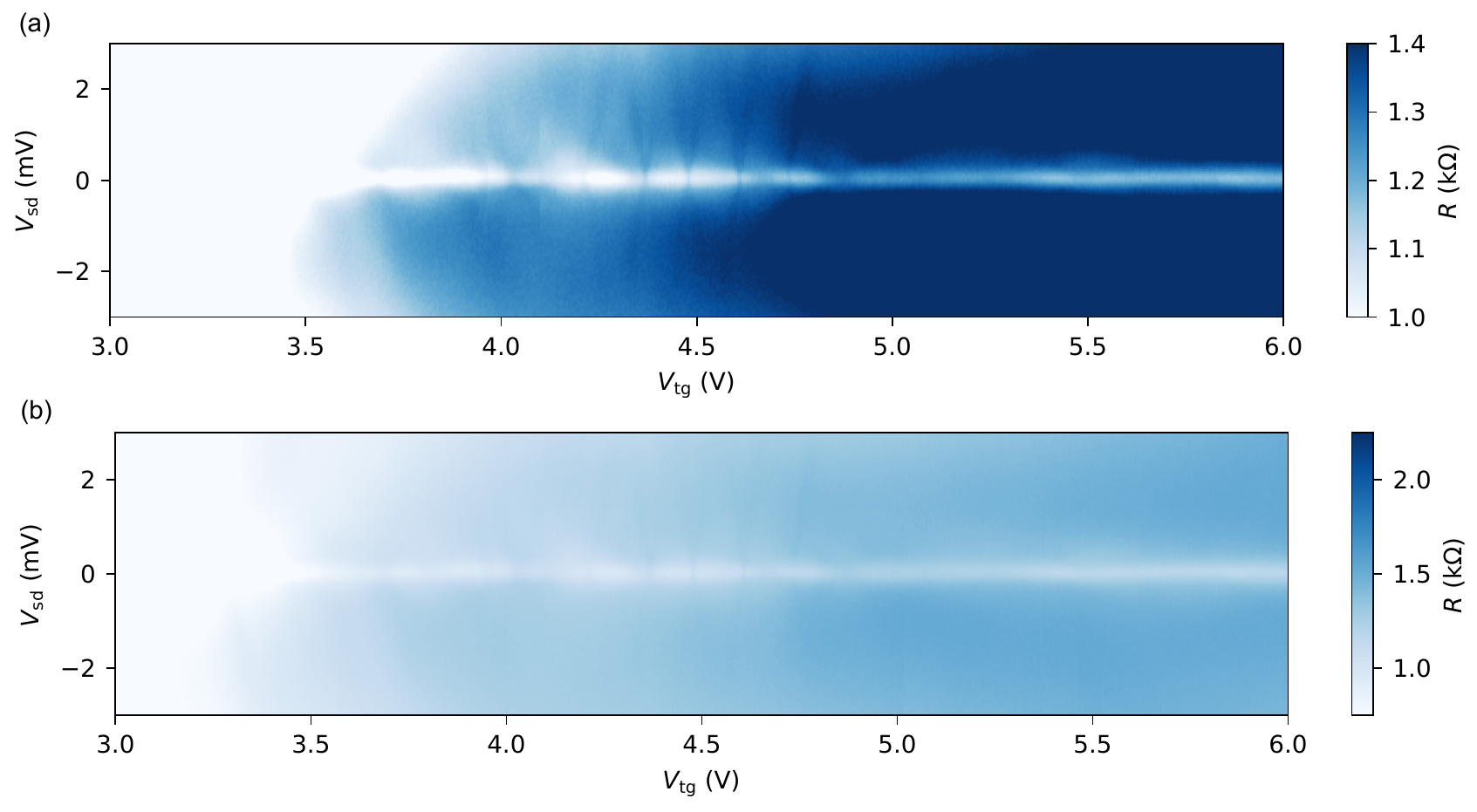}
\caption{(a) Conductance as a function of source-drain voltage and top gate voltage with an extended range in top gate. We observe Coulomb diamonds which fade away for top gate values outside of the $\SI{4}{V} < V_{\mathrm{tg}} < \SI{5}{V}$ range.}
\label{fig:S4}
\end{figure*}

Figure \ref{fig:S4} shows the extended range measurement from the data shown in Fig. 3(c). Figure \ref{fig:S4}(a) keeps the same color scale as in the main text, which leads to a better visibility of the features in the central region but saturates at high and low source-drain voltages. Fig. \ref{fig:S4}(b) has an increased range for the color scale, avoiding the saturation in most of the figure but loosing visibility in the central region.